\documentclass[usenatbib]{mn2e}
\usepackage{graphicx,times}
\usepackage{bm,url}
\graphicspath{{./fig/}{./png/}}
\newcommand{\EQ}{\begin{equation}}
\newcommand{\EN}{\end{equation}}
\newcommand{\EQA}{\begin{eqnarray}}
\newcommand{\ENA}{\end{eqnarray}}
\newcommand{\nab}{\bm{\nabla}}
\newcommand{\kf}{k_{\rm f}}

\newcommand{\meanEMF}{\overline{\mbox{\boldmath ${\mathcal E}$}} {}}

\newcommand{\meanB}{\overline{{B}}}
\newcommand{\meanJ}{\overline{{J}}}
\newcommand{\meanBB}{\bm{\overline{{B}}}}
\newcommand{\meanJJ}{\bm{\overline{{J}}}}
\newcommand{\meanUU}{\bm{\overline{{U}}}}

\newcommand{\uu}{\bm{{{u}}}}
\newcommand{\UU}{\bm{{{U}}}}
\newcommand{\bb}{\bm{{{b}}}}
\newcommand{\BB}{\bm{{{B}}}}
\newcommand{\JJ}{\bm{{{J}}}}
\newcommand{\ff}{\bm{{{f}}}}

\newcommand{\FF}{\bm{{{F}}}}

\newcommand{\dd}{\mbox{d}}
\newcommand{\Bhat}{\hat{B}}
\newcommand{\BBhat}{\hat{\bm{B}}}
\newcommand{\zzz}{\hat{\bm{z}}}

\def\const{{\rm const}}

\def\Rm{R_\mathrm{m}}
\def\Rey{\mbox{\rm Re}}

\def\half{{\textstyle{1\over2}}}

\def\onethird{{\textstyle{1\over3}}}

\newcommand{\SSSS}{\mbox{\boldmath ${\sf S}$} {}}

\newcommand{\EEq}[1]{Equation~(\ref{#1})}
\newcommand{\Eq}[1]{equation~(\ref{#1})}

\newcommand{\Eqss}[2]{equations~(\ref{#1})--(\ref{#2})}

\newcommand{\Sec}[1]{Sect.~\ref{#1}}

\newcommand{\Fig}[1]{Fig.~\ref{#1}}

\title[A growing dynamo from a saturated Roberts flow dynamo]
{A growing dynamo from a saturated Roberts flow dynamo}
\author[A. Tilgner \& A.~Brandenburg]%
{Andreas Tilgner$^1$ and Axel Brandenburg$^2$\\
$^1$ Institute of Geophysics, University of G\"ottingen, Friedrich-Hund-Platz 1, 37077 G\"ottingen, Germany \\
$^2$ NORDITA, Roslagstullsbacken 23, SE - 106 91 Stockholm, Sweden
}

\date{}

\begin{document}

\pagerange{\pageref{firstpage}--\pageref{lastpage}} \pubyear{2008}

\maketitle

\begin{abstract}
Using direct simulations, weakly nonlinear theory and nonlinear mean-field
theory, it is shown that the quenched velocity field of a saturated
nonlinear dynamo can itself act as a kinematic dynamo.
The flow is driven by a forcing function that would produce a Roberts flow
in the absence of a magnetic field.
This result confirms an analogous finding by F.\ Cattaneo \& S.\ M.\ Tobias
(arXiv:0809.1801) for the more complicated case
of turbulent convection, suggesting that this may be a common property
of nonlinear dynamos; see also the talk given also online at the
Kavli Institute for Theoretical Physics
(http://online.kitp.ucsb.edu/online/dynamo\_c08/cattaneo).
It is argued that this property can be used to test nonlinear mean-field
dynamo theories.
\end{abstract}
\label{firstpage}
\begin{keywords}
magnetic fields --- MHD --- hydrodynamics
\end{keywords}

\section{Introduction}

The magnetic fields of many astrophysical bodies displays order on scales
large compared with the scale of the turbulent fluid motions that are
believed to generate these fields via dynamo action.
A leading theory for these types of dynamos is mean-field electrodynamics
\citep{Mof78,KR80}, which predicts the evolution of suitably averaged
mean magnetic fields.
Central to this theory is the mean electromotive force based on the
fluctuations of velocity and magnetic fields.
This mean electromotive force is then expressed in terms of the
mean magnetic field and its first derivative with coefficients
$\alpha_{ij}$ and $\eta_{ijk}$.
The former represents the $\alpha$ effect and the latter the turbulent
magnetic diffusivity.

Under certain restrictions the coefficients $\alpha_{ij}$ and
$\eta_{ijk}$ can be calculated using for example the first order smoothing
approximation, which means that nonlinearities in the evolution equations
for the fluctuations are neglected.
Whilst this is a valid approach for small magnetic Reynolds numbers or
short correlation times,
it is not well justified in the astrophysically interesting
case when the magnetic Reynolds number is large and the correlation time
comparable with the turnover time.
However, in recent years it has become possible to calculate $\alpha_{ij}$
and $\eta_{ijk}$ using the so-called test-field method \citep{S05,S07}.
For the purpose of this paper we can consider this method essentially
as a ``black box'' whose input is the velocity field and its output are
the coefficients $\alpha_{ij}$ and $\eta_{ijk}$.
This method has been successfully applied to the kinematic case of weak
magnetic fields in the presence of homogeneous turbulence either without
shear \citep{Sur_etal08,Betal08b} or with shear \citep{B05,Betal08}.

More recently, this method has also been applied to the nonlinear case
where the velocity field is modified by the Lorentz force associated
with the dynamo-generated field \citep{Betal08c}
In that case the test-field method consists still of the same black box,
whose input is only the velocity field, but now this velocity field is
based on a solution of the full hydromagnetic equations comprising the
continuity, momentum, and induction equations.
We emphasize that the magnetic field is quite independent of the fields
that appear in the test-field method inside the black box.

Our present work is stimulated by an interesting and relevant numerical
experiment performed recently by \cite{CT08}.
They considered a solution of the full hydromagnetic equations where the
magnetic field is generated by turbulent convective dynamo action and
has saturated at a statistically steady value.
They then used this velocity field and subjected it to an independent
induction equation, which is equal to the original induction equation
except that the magnetic field $\BB$ is now replaced by a passive vector
field $\tilde{\BB}$, which does not react back on the momentum equation.
Surprisingly, they found that $|\tilde{\BB}|$ grows
exponentially, even though the velocity field is already quenched by
the original magnetic field.

One might have expected that, because the velocity is modified
such that it produces a statistically steady solution to the original
induction equation, $\tilde{\BB}$ should decay or
also display statistically steady behavior.
The argument sounds particularly convincing for time-independent flows because,
if a growing $\tilde{\BB}$ were to exist, one would expect this alternative field to grow and
replace the initial field. This view is supported by recent simulations in which
the flow field from a geodynamo simulation in a spherical shell was used as
velocity field in kinematic dynamo computations, and no growing $\tilde{\BB}$
was found \citep{Tilgner08}.
However, it turns out that this reasoning is not correct in general.
One finds counterexamples even within the confines of
mean field MHD using analytical tools. The existence of a growing $\tilde{\BB}$
thus is not tied to chaotic flows or fluctuating small-scale dynamos.

This finding of \cite{CT08} is interesting in view of the applicability
of the test-field method to the nonlinear case.
Of course, the equations used in the test-field method are different from
the original induction equation.
(The equations used in the test-field method include an inhomogeneous term and
the mean electromotive force is subtracted, but they are otherwise
similar to the original induction equation.)
Given the seemingly unphysical behaviour of the induction equation in the
presence of a vector field different from the actual magnetic field, it would
be tempting to argue that one should choose test fields whose shape is
rather close to that of the actual magnetic field \citep{CH08}.
On the other hand, the $\alpha_{ij}$ and $\eta_{ijk}$ tensors should
give the correct response to all possible fields, not just the $\BB$ field that
grew out of a particular initial condition, but also the passive
$\tilde{\BB}$ field that obeys a separate induction equation.
It is therefore important to choose a set of test fields that are orthogonal to
each other, even if none of the fields are solutions of the induction equation.
One goal of this paper is to show that the $\alpha_{ij}$
and $\eta_{ijk}$ tensors obtained in this way provide not only interesting
diagnostics of the flow, but they are also able to explain
the surprising result of \cite{CT08} in the context of a simpler example.
However, let us begin by repeating the numerical experiment of \cite{CT08}
using the simpler case of a Roberts flow.
Next, we consider a weakly nonlinear analysis of this problem and turn
then to its mean-field description.

\section{The model}

In order to examine the possibility of a growing passive vector field,
we first considered the case of a driven ABC flow.
Such a flow is non-integrable and has chaotic streamlines. Growing
passive vector fields were found. To simplify matters even further,
we consider now the case of a Roberts flow,
which is integrable, has non-chaotic streamlines, and the dynamo can only
be a slow one, i.e.\ the growth rate goes to zero in the limit of large
magnetic Reynolds number.
This is however not an issue here, because we will only be considering
finite values of the magnetic Reynolds number.

In the following we consider both incompressible and isothermal cases.
The governing equations for any externally driven velocity field
(turbulence, ABC flow, or Roberts flow) are then given by
\EQ
{\partial\UU\over\partial t}=-\UU\cdot\nab\UU-\nab H+{1\over\rho}\JJ\times\BB
+\ff+\FF_{\rm visc},
\label{dUdt}
\EN
\EQ
{\partial\BB\over\partial t}=\nab\times(\UU\times\BB)+\eta\nabla^2\BB,
\EN
where $\UU$ is the velocity, $\BB$ is the magnetic field, $\rho$ is
the density, $H$ is the specific enthalpy, $\JJ=\nab\times\BB/\mu_0$ is
the current density, $\mu_0$ is the vacuum permeability, $\ff$ is the
forcing function, $\FF_{\rm visc}$ is the viscous force per unit mass,
and $\eta=\const$ is the magnetic diffusivity.
In the incompressible case, $\nab\cdot\UU=0$, we have $H=p/\rho$, where
$p$ is the pressure and $\rho=\const$.
The viscous force is then given by $\FF_{\rm visc}=\nu\nabla^2\UU$.
In the isothermal case, the density obeys the usual continuity equation
\EQ
{\partial\rho\over\partial t}=-\nab\cdot(\rho\UU),
\EN
but now $p=c_{\rm s}^2\rho$, where $c_{\rm s}$ is the isothermal sound speed.
In that case $H=c_{\rm s}^2\ln\rho$ and the viscous force is given by
\EQ
\FF_{\rm visc}=\nu\nabla^2\UU+\onethird\nu\nab\nab\cdot\UU
+2\nu\SSSS\nab\ln(\rho\nu),
\EN
where ${\sf S}_{ij}=\half(U_{i,j}+U_{j,i})-\onethird\delta_{ij}\nab\cdot\UU$
is the traceless rate of strain matrix.

In order to compute the evolution of an additional passive vector field
$\tilde{\BB}$ we also solve the equation
\EQ
{\partial\tilde{\BB}\over\partial t}=\nab\times(\UU\times\tilde{\BB})
+\eta\nabla^2\tilde{\BB}.
\label{dBtildedt}
\EN
In the case of the Roberts flow we use the forcing function
\EQ
\ff=\nu\kf^2\UU_{\rm Rob},
\EN
where
\EQ
\UU_{\rm Rob}=\kf\psi\zzz-\zzz\times\nab\psi
\EN
with
\EQ
\psi=(u_0/k_0)\cos k_0x\cos k_0y
\EN
and $k_{\rm f}=\sqrt{2}k_0$.
We consider a domain of size $L_x\times L_y\times L_z$.
In all cases we consider $L_x=L_y=L_z=2\pi/k_0$.
Our model is characterized by the choice of fluid and magnetic Reynolds
numbers that are here based on the inverse wavenumber $k_0$,
\EQ
\Rey=u_0/\nu k_0,\quad
\Rm=u_0/\eta k_0.
\EN

\section{Numerical experiments}
\label{NumericalExperiments}

\begin{figure}\begin{center}
\includegraphics[width=\columnwidth]{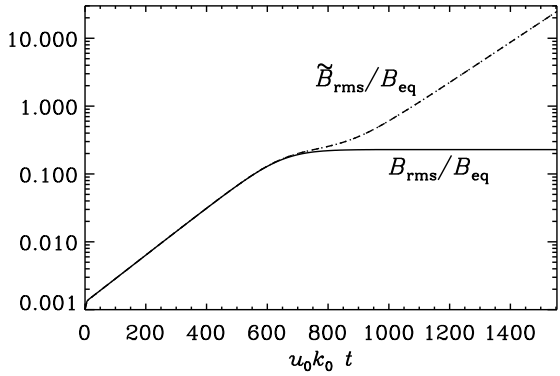}
\end{center}\caption[]{
Evolution of the rms values of $\BB$ and $\tilde{\BB}$ for $\Rm=6.25$.
The growth rate of $\tilde{\BB}$ is $0.016\,u_0k_0$ in the kinematic phase
and $0.014\,u_0k_0$ in the nonlinear phase.
}\label{p_32b}\end{figure}

We solve \Eqss{dUdt}{dBtildedt} for the isothermal case
using the {\sc Pencil Code}\footnote{%
\url{http://www.nordita.org/software/pencil-code/}}, which is a high-order
public domain code (sixth order in space and third order in time) for
solving partial differential equations.
\EEq{dBtildedt} is solved using the test-field module with the input
parameters \texttt{lignore\_uxbtestm=T}, and \texttt{itestfield='B=0'},
which means that the inhomogeneous term of the test-field equation is
set to zero and the subtraction of the mean electromotive force has
been disabled.
In this way we solve \Eq{dBtildedt}, instead of the original
test-field equation.
We focus on the case of small fluid Reynolds number, $\Rey=0.5$.
The initial conditions for $\BB$ and $\tilde{\BB}$ are Beltrami fields,
$(\cos(k_0z+\varphi),\sin(k_0z+\varphi), 0)$,
with an arbitrarily chosen phase $\varphi=0.2$, but for $\tilde{\BB}$
we $\varphi=0$.
In \Fig{p_32b} we plot the evolution of the rms values of $\BB$ and
$\tilde{\BB}$ for a weakly supercritical case with $\Rm=6.25$.
(In our case with $\Rey=0.5$ the critical value for dynamo action is
$\Rm\approx5.77$; for $\Rey\to0$ the critical value would be $\Rm\approx5.5$.)
Both $\BB$ and $\tilde{\BB}$ grow at first exponentially at the same rate.
However, when $\BB$ reaches saturation, the growth of $\tilde{\BB}$ slows
down temporarily, but then resumes to nearly its original value. This
confirms the result of \cite{CT08} for the much simpler case of a Roberts flow.

\begin{figure}\begin{center}
\includegraphics[width=\columnwidth]{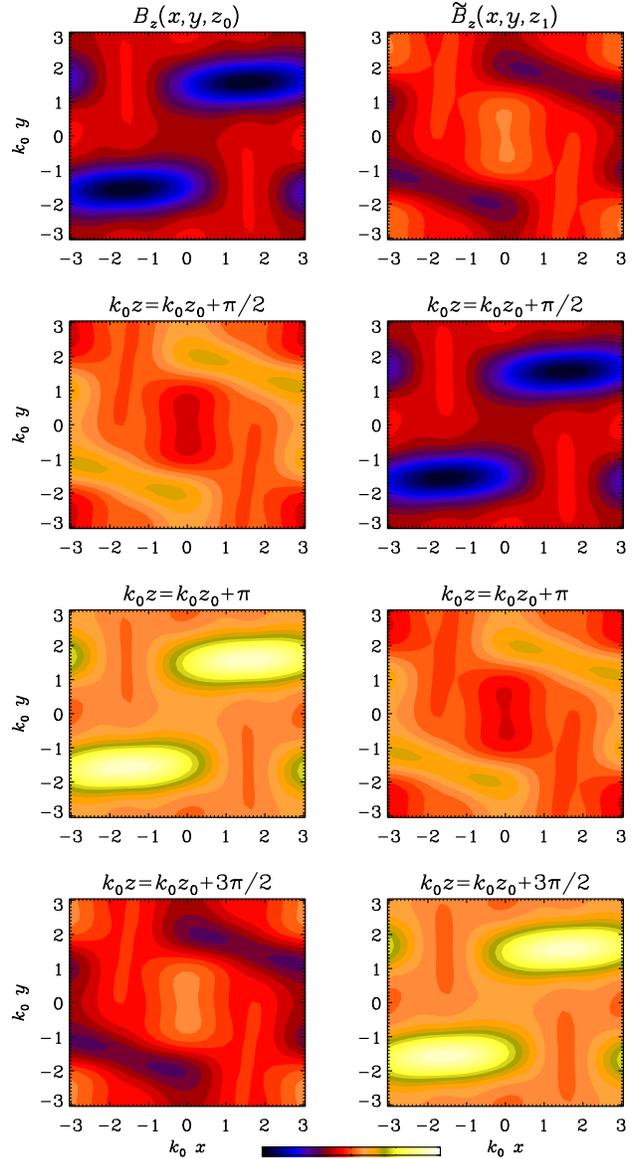}
\end{center}\caption[]{
Grey-scale representations of
horizontal cross-sections of $\BB$ and $\tilde{\BB}$ for $\Rm=6.25$.
Here, $k_0z_0\approx-3.04$.
Both fields are scaled symmetrically around zero (grey shade)
with dark shades indicating negative values and light shades
indicating positive vales, as indicated on the greyscale bar.
}\label{RobertsForced_32b_pxy}\end{figure}

In \Fig{RobertsForced_32b_pxy} we compare horizontal cross-sections of
the two fields.
Note that the two are phase shifted in the $z$ direction by a quarter wavelength.
The short interval in \Fig{p_32b} during which the growth of $\tilde{\BB}$
slowed down temporarily is therefore related to the fact that the solution
needed to ``adjust'' to this particular form.

\section{Weakly nonlinear theory}

A weakly nonlinear analysis of the Roberts flow is presented in
\cite{Tilgner01}. Two nonlinear terms enter the full dynamo problem. In order to
make the calculation analytically tractable, it is assumed that the fluid has
infinite magnetic Prandtl number so that the inertial terms (and hence the
advection term) drop from the Navier-Stokes equation. The second nonlinear term,
the Lorentz force, is assumed to be small compared with the driving force $\ff$
and is treated perturbatively. The linear induction equation is transformed into
a mean field equation assuming separation of length scales and small magnetic Reynolds
numbers. One can under these assumptions compute the modifications of the velocity field
$\UU_{\rm Rob}$ due to the presence of a mean field $\meanBB$,
which we define here as
\EQ
\meanBB(z,t)=\int_0^{L_y}\int_0^{L_x}\BB(x,y,z,t)\,\dd x\,\dd y/L_x L_y.
\label{meanBBdef}
\EN
The magnetically modified velocity field then becomes
\EQ
\UU=(1-\gamma) \UU_{\rm Rob} + 2 \gamma \frac{\meanB_x \meanB_y}{\meanB_x^2 + \meanB_y^2} \hat{\UU}
\label{flowU}
\EN
with
\EQ
\hat{\UU}=u_0 \pmatrix {\sin k_0 x \cos k_0 y \cr
- \cos k_0 x \sin k_0 y \cr \sqrt{2} \sin k_0 x \sin k_0 y}
\EN
and $\gamma=(\meanB_x^2+\meanB_y^2)/(2 \eta \nu k_{\rm f}^2 \rho \mu_0)$.
The original flow $\UU_{\rm Rob}$ is reduced in magnitude and another 2D periodic
flow component is added. The mean-field induction equation with flow $\UU$ given
by (\ref{flowU}), written for a passive vector field $\tilde{\BB}$, is:
\EQ
\frac{\partial \tilde{\BB}}{\partial t}
+ \nab \times A \pmatrix{\tilde{B}_x \cr \tilde{B}_y \cr 0} 
- \nab \times C \meanB_x \meanB_y
\pmatrix{\tilde{B}_y \cr \tilde{B}_x \cr 0}
= \eta \nab^2 \tilde{\BB}
\label{quench}
\EN
with
\EQ
A=\tilde{R}_{\rm m} v_0\quad\mbox{and}\quad
C={\tilde{R}_{\rm m}^3\over P_{\rm m}} {1\over v_0 \rho \mu_0},
\EN
where
\EQ
\tilde{R}_{\rm m}={v_0\over\eta\kf},\quad
 P_{\rm m}={\nu\over\eta},\quad\mbox{and}\quad
v_0^2=\half u_0^2 (1-\gamma)
\EN
have been introduced.
This equation corresponds to equation~(10) of \cite{Tilgner01}. Apart from a
change of notation, the distinction between the passive vector field
$\tilde{\BB}$ and the field distorting the Roberts flow, $\meanBB$, has been
made. In addition, equation~(10) of \cite{Tilgner01} was intended as a model of
the Karls\-ru\-he dynamo and used $u_0$ as control parameter, whereas here, we
consider $\ff$ as given. For this reason, the two equations are identical only
to first order in $\gamma$.

Equation~(\ref{quench}) reduces to the usual dynamo problem for
$\tilde{\BB}=\meanBB$ and leads to the kinematic dynamo problem if it is furthermore
linearized in $\meanBB$, which corresponds to dropping the third term
in Eq.~(\ref{quench}). For $u_0^2 > 2 \eta^2 k_{\rm f} k$, the equation has
growing solutions of the form $(\cos kz, \sin kz, 0)$. Weakly nonlinear analysis
determines the amplitude $B_0$ of a saturated solution of the form
$\meanBB = B_0 (\cos kz, \sin kz, 0)$ by inserting this ansatz for $\meanBB$ and
$\tilde{\BB}=\meanBB$ into equation~(\ref{quench}) and by projecting
equation~(\ref{quench})
onto $(\cos kz, \sin kz, 0)$ and integrating spatially over a periodicity
cell \citep{Tilgner01}.

We proceed similarly to find growing solutions of equation~(\ref{quench}). Assume $\meanBB = B_0 (\cos
kz, \sin kz, 0)$. Obvious candidates for passive vector fields growing at rate $p$ are
$\tilde{\BB} \propto e^{pt} (\cos (kz+\varphi), \sin (kz+\varphi), 0)$. Since
the third term in equation~(\ref{quench}) is a perturbation, a growing solution
must be of a form such that the other terms maximize the growth rate. These other
terms are identical to the kinematic problem, so that $\tilde{\BB}$
must have the same general form as the kinematic dynamo field except for the
phase shift $\varphi$ which measures the phase angle between the saturated field
$\meanBB$ and the passive vector field $\tilde{\BB}$. The velocity field $\UU_{\rm Rob}$ is
independent of $z$, so that any solution of the kinematic dynamo problem remains
a solution after translation along $z$. However, neither the Lorentz force due
to $\meanBB$ nor the flow modified by that Lorentz force are independent of $z$,
so that the phase angle $\varphi$ matters for $\tilde{\BB}$. The above ansatz for
$\tilde{\BB}$ will not be an exact solution of equation~(\ref{quench}) because of the
$z$-dependence of $\meanB_x \meanB_y$, but it represents the leading Fourier
component. In order to determine the optimal $\varphi$, we insert this ansatz
into equation~(\ref{quench}), project onto $(\cos (kz+\varphi), \sin (kz+\varphi),
0)$ and integrate over a periodicity cell to find
\EQ
p = -\eta k^2 + A k - C B_0^2 k \frac{1}{4} \cos 2\varphi.
\EN
The saturation amplitude $B_0$ is determined from this equation by setting $p=0$
and $\varphi=0$. For any given $B_0$, the maximum of $p$ is obtained for
$\varphi =\pi/2$. The fastest growing mean passive vector field is thus expected to have
the same form as the mean dynamo field except for a translation by a quarter
wavelength along the $z$-axis.
This is in agreement with the simulation results of \Sec{NumericalExperiments}.

The weakly nonlinear analysis in summary delineates the following physical
picture: As detailed in \cite{Tilgner01}, the saturated dynamo field modifies
the flow in two different ways. Firstly, it reduces the amplitude of $\UU_{\rm Rob}$
by the factor $1-\gamma$, and
secondly, it introduces a new set of vortices which lead to the third term in
equation~(\ref{quench}). The reduction of the amplitude of the Roberts flow affects
all magnetic mean fields with a spatial dependence in $(\cos (kz+\varphi), \sin
(kz+\varphi), 0)$ in the same manner, independently of $\varphi$. The additional
vortices, however, have a quenching effect on the field that created them (e.g.
$\varphi=0$) but are amplifying for a field shifted by $\varphi = \pi/2$ with
respect to the saturated field.

We were able to find a simple growing passive vector field thanks to the periodic boundary
conditions in $z$. The same construction is impossible for vacuum boundaries at
$z=0$ and $z=2 \pi/k_0$. Numerical simulations of the Roberts dynamo with vacuum
boundaries, not reported in detail here, have revealed that growing
$\tilde{\BB}$ exist in this geometry nonetheless, but they bear a more
complicated relation with $\BB$ than a simple translation. At present, the flow
of the convection driven dynamo in a spherical shell used in \cite{Tilgner08}
seems to be the only known example of a dynamo which does not allow for growing
$\tilde{\BB}$.

\section{Nonlinear mean-field theory}

In mean-field dynamo theory for a flow such as the Roberts flow
one solves an equation for the horizontally averaged mean field,
as defined in \Eq{meanBBdef}.
The mean electromotive force is defined as $\meanEMF=\overline{\uu\times\bb}$,
where $\uu=\UU-\meanUU$ and $\bb=\BB-\meanBB$ are the fluctuating components
of magnetic and velocity fields.
The mean electromotive force can be expressed in terms of the mean fields as
\EQ
{\cal E}_i=\alpha_{ij}\meanB_j-\eta_{ij}\meanJ_j,
\EN
where we have used the fact that for mean fields that depend only on
one spatial coordinate one can express all first derivatives of the components
of $\meanBB$ in terms of those of $\meanJJ$ alone.

The tensorial forms of $\alpha_{ij}$ and $\eta_{ij}$ are ignored
in many mean-field dynamo applications,
but here their tensorial forms turn out to be of crucial importance.
Of course, there is always the anisotropy with respect to the $z$
direction, but this is unimportant in our one-dimensional mean field
problem, because of solenoidality of $\meanBB$ and $\meanJJ$ and suitable
initial conditions on $\meanBB$ such that $\meanB_z=\meanJ_z=0$.
However, the dynamo-generated magnetic fields will introduce an anisotropy
in the $x$ and $y$ directions.
If $\meanBB$ is the only vector giving a preferred direction
to the system, the $\alpha_{ij}$ and $\eta_{ij}$ tensors must be of the form
\EQ
\alpha_{ij}=\alpha_1(\meanBB)\delta_{ij}+\alpha_2(\meanBB)\Bhat_i\Bhat_j+...,
\label{alpha}
\EN
\EQ
\eta_{ij}=\eta_1(\meanBB)\delta_{ij}+\eta_2(\meanBB)\Bhat_i\Bhat_j+...,
\EN
where $\BBhat=\meanBB/|\meanBB|$ is the unit vector of the dynamo-generated
mean magnetic field, and dots indicate the presence of terms
related to the anisotropy in the $z$ direction inherent to the Roberts flow.
As indicated above, equation (\ref{alpha}) is correct without these terms
only in the $(x,y)$ plane.
However, the terms represented by the dots do not enter the considerations
below because we are only interested in fields with $\meanB_z=0$.

In order to predict the evolution of $\tilde{\BB}$ in the saturated state,
we need to know the effect on $\alpha_{ij}$ (and in principle also on
$\eta_{ij}$, but $\eta_2$ is small; see below).
Thus, we now need to know $\meanBB$.
The mean magnetic field generated by the Roberts flow is a force-free
Beltrami field of the form
\EQ
\meanBB=(\cos k_0z,\sin k_0z, 0),
\EN
so
\EQ
\Bhat_i\Bhat_j=\pmatrix{
\cos^2 k_0z & \cos k_0z\sin k_0z & 0\cr
\cos k_0z\sin k_0z & \sin^2 k_0z & 0\cr
0 & 0 & 0}.
\EN
The coefficients $\alpha_1$, $\alpha_2$, $\eta_1$, and $\eta_2$ have
previously been determined for the case of homogeneous turbulence
\citep{Betal08c} and it turned out at $\alpha_1$ and $\alpha_2$ have
opposite sign, and that $\eta_2$ is negligible.
This is also true in the present case, for which we have determined
$\alpha_1/u_0=-0.266$, $\alpha_2/u_0=+0.022$,
$\eta_1 k_0/u_0=0.082$, and $\eta_2 k_0/u_0=0.002$.
The microscopic value of $\eta$ is 0.160, so the steady state condition,
$\alpha_1+\alpha_2+(\eta_1+\eta_2+\eta)k_0=0$, is obeyed.\footnote{%
We note that the sign of $\alpha_1$ is opposite to the sign of
the kinetic helicity, but since the Roberts flow has positive
helicity, $\alpha_1$ must be negative, which is indeed the case.}
In the kinematic regime we have $\alpha_1/u_0=-0.254$, $\eta_1 k_0/u_0=0.076$,
with $\alpha_2=\eta_2=0$, resulting in a positive growth rate of
$0.018\,u_0k_0$.
Thus, even though $\alpha_1$ increases in this case,
the sum $\alpha_1+\alpha_2$ is being quenched.
This, together with the increase of $\eta_1+\eta_2$, leads to saturation
of $\meanBB$.

Returning now to the mean-field problem for $\tilde{\BB}$, this too
will be governed by the same $\alpha_{ij}$ and $\eta_{ij}$ tensors, but
now the tensor $\Bhat_i\Bhat_j$ is fixed and independent of $\tilde{\BB}$.
The solution for $\tilde{\BB}$ will be one that maximizes the
growth, so it must experience minimal quenching.
Such a solution is given by that eigenvector of $\Bhat_i\Bhat_j$
that minimizes the quenching of $\tilde{\BB}$.
In the case of our Beltrami field (\ref{alpha}), the minimizing eigenvector is given by
\EQ
\tilde{\meanBB}=(\sin k_0z,-\cos k_0z, 0),
\EN
which satisfies $\Bhat_i\Bhat_j\tilde{\meanB}_j=0$.
This is indeed the same result that we found both numerically and
using weakly nonlinear theory.
The growth rate of $\tilde{\BB}$ is then expected to be
$|\alpha_1|k_0-(\eta_1+\eta)k_0^2=0.024\,u_0k_0$, which is indeed positive,
but it is somewhat bigger than the one seen in \Fig{p_32b}.

Let us emphasize once more that by determining the full
$\alpha_{ij}$ and $\eta_{ij}$ tensors in the nonlinear case,
we have been able to predict the behavior of the passive vector
field as well.
This adds to the credence of the test-field method in the nonlinear case,
and confirms that the test-fields can well be very different from the
actual solution.

The considerations above suggest that solutions to the passive vector
equation \Eq{dBtildedt} can be used to provide an independent test
of proposed forms of $\alpha$ quenching.
Isotropic formulations of $\alpha$ quenching would not
reproduce the growth of a passive vector field, and so such quenching
expressions can be ruled out, even though the resulting electromotive
force for $\meanBB$ would be the same.
We suggest therefore that the eigenvalues and eigenvectors of
\Eq{dBtildedt} with a velocity field from a saturated dynamo can
be used to characterize the quenching of dynamo parameters
($\alpha$ effect and turbulent diffusivity) and thereby to test
proposed forms of $\alpha$ quenching.

\section{Conclusions}

The most fundamental question of dynamo theory beyond kinematic dynamo theory is
``how do magnetic fields saturate?''. In the simplest picture, the velocity
field reorganizes in response to the Lorentz force such that all magnetic fields
decay except one which has zero growth rate and which is the one we observe.
This picture is already questionable for chaotic dynamos. In a chaotic system,
nearby initial conditions lead to exponentially separating time evolutions. If
one takes a magnetic field $\BB$ with (on time average) zero growth rate which
is the saturated solution of a chaotic dynamo, and solves the kinematic dynamo
problem for a passive vector field $\tilde{\BB}$ with initial conditions
different from $\BB$, one is prepared to find growing $\tilde{\BB}$. 
Examples for growing $\tilde{\BB}$ in chaotic dynamos have been given by
\cite{CT08}.

For a time-independent saturated dynamo,
on the other hand, the simple picture seems to
be adequate at first. However, we have shown in this paper that growing
$\tilde{\BB}$ also exist in the time-independent Roberts dynamo. The origin of
the growing $\tilde{\BB}$ can in this case be understood with the help of weakly
nonlinear theory. The growing $\tilde{\BB}$ has the same shape as the saturated
dynamo field but is translated in space. 

What was wrong with the naive intuition invoked above? It was based on a
stability argument (the equilibrated magnetic field should be replaced by
$\tilde{\BB}$ if there is a growing $\tilde{\BB}$). However, the linear
stability problem for a solution of the full dynamo equations is different from
the kinematic dynamo problem for $\tilde{\BB}$, because in the latter, the
velocity field is fixed. Both problems are closely related eigenvalue problems, but standard
mathematical theorems do not provide us with a relation between the spectra of
both problems. The numerical computation above gives an example of a stable
Roberts dynamo, showing that the linear stability problem for the solution found
there has only negative eigenvalues. Solving the same eigenvalue problem with
velocity fixed, which is the kinematic problem for $\tilde{\BB}$, can very well
lead to positive eigenvalues, and indeed, it does. The dynamo is therefore only
stable because the velocity field is able to adjust to perturbations in the
magnetic field. The magnetic field on its own is unstable.

\section*{Acknowledgments}

We acknowledge the Kavli Institute for Theoretical Physics
for providing a stimulating atmosphere during the program on dynamo theory.
This work has been initiated through discussions with
Steve M.\ Tobias and Fausto Cattaneo.
We thank Eric G.\ Blackman, K.-H.\ R\"adler, and Kandaswamy Subramanian
for comments on the paper.
This research was supported in part by the National Science
Foundation under grant PHY05-51164.

\newcommand{\ybook}[3]{ #1, {#2} (#3)}
\newcommand{\yjfm}[3]{ #1, {J.\ Fluid Mech.,} {#2}, #3}
\newcommand{\yprl}[3]{ #1, {Phys.\ Rev.\ Lett.,} {#2}, #3}
\newcommand{\ypre}[3]{ #1, {Phys.\ Rev.\ E,} {#2}, #3}
\newcommand{\yapj}[3]{ #1, {ApJ,} {#2}, #3}
\newcommand{\ynat}[3]{ #1, {Nat} {#2}, #3}
\newcommand{\yptrsa}[3]{ #1, {Phil. Trans. Roy. Soc. London A} {#2}, #3}
\newcommand{\ymn}[3]{ #1, {MNRAS,} {#2}, #3}
\newcommand{\yan}[3]{ #1, {AN,} {#2}, #3}
\newcommand{\yana}[3]{ #1, {A\&A,} {#2}, #3}
\newcommand{\ygafd}[3]{ #1, {Geophys.\ Astrophys.\ Fluid Dyn.,} {#2}, #3}
\newcommand{\ypf}[3]{ #1, {Phys.\ Fluids,} {#2}, #3}
\newcommand{\yproc}[5]{ #1, in {#3}, ed.\ #4 (#5), #2}
\newcommand{\yjour}[4]{ #1, {#2} {#3}, #4.}
\newcommand{\sapj}[1]{ #1, {ApJ,} submitted}
\newcommand{\papj}[1]{ #1, {ApJ,} in press}
\newcommand{\smn}[1]{ #1, {MNRAS,} submitted}
\newcommand{\pmn}[1]{ #1, {MNRAS,} in press}


\label{lastpage}
\end{document}